\documentclass[prb,twocolumn,showpacs,floatfix]{revtex4}
\usepackage{graphicx}
\usepackage{dcolumn}
\usepackage{bm}
\begin{document}
\title{Hyperfine interaction in CoCl$_2$  investigated by high resolution neutron spectroscopy}
\author{Tapan Chatterji$^1$ and Bernhard Frick$^1$}
\address{$^1$Institut Laue-Langevin, BP 156, 38042 Grenoble Cedex 9, France\\
}
\date{\today}

\begin{abstract}
We investigated low energy nuclear spin excitations in the layered compound CoCl$_2$ by high resolution back-scattering neutron spectroscopy. We detected inelastic peaks at $E = 1.34 \pm 0.03$ $\mu$eV on both energy loss and energy sides of the central elastic peak at $T = 2$ K. The energy of the inelastic peaks decrease with temperature continuously and become zero at $T_N \approx 25$ K at which the two ielastic peaks merge with the central elastic peak. We interpret the low energy excitations to be due to the transition between hyperfine field split nuclear levels. The present data together with the data on other Co compounds show that the energy of the nuclear spin excitations of a number of compounds follow a linear relationship with the electronic magnetic moment of the Co ion whereas that of other compounds deviate appreciably from this linear behaviour. We ascribe this anomalous behaviour to the presence of unquenched orbital moments of the Co ions.

\end{abstract}
\maketitle

The hyperfine field of an atom or ion is the magnetic field at the atomic nucleus produced by the electrons in the solid due to the hyperfine interaction between the magnetic moment of the electrons and that of the nucleus \cite{freeman65}. This interaction can be measured by the M\"ossbauer effect or by the nuclear magnetic resonance (NMR) technique. Another less well-known method is the spin-flip scattering of neutrons measured by high resolution neutron spectroscopy \cite{heidemann70}. It is well-known that the magnetic hyperfine fields in a solid gives valuable information about the electronic structure and the magnetic properties of the solid. The hyperfine field is a valuable probe of electron spin density distribution at the nuclei. It can sometimes be related to the electronic magnetic moment.  The hyperfine field is site and element selective. The hyperfine field $B_{hf}$ can be given by
\begin{equation}
B_{hf}=B^s_{hf}+B^d_{hf}+B^o_{hf}
\end{equation}
where $B^s_{hf}$ is the Fermi contact term due to the s electrons, $B^d_{hf}$ is magnetic dipole and $B^o_{hf}$ is the orbital term due to the non-s electrons. Normally the Fermi contact term is the most dominant term whereas the magnetic dipole term is often very small and can be neglected. The orbital term is appreciable for rare earth ions except for Eu$^{2+}$ and Gd$^{3+}$, which have no orbital moment. The orbital moment is usually quenched in some 3d Fe series elements. However in some compounds of Co and V it can be quite important.
 
\begin{figure}
\resizebox{0.35\textwidth}{!}{\includegraphics{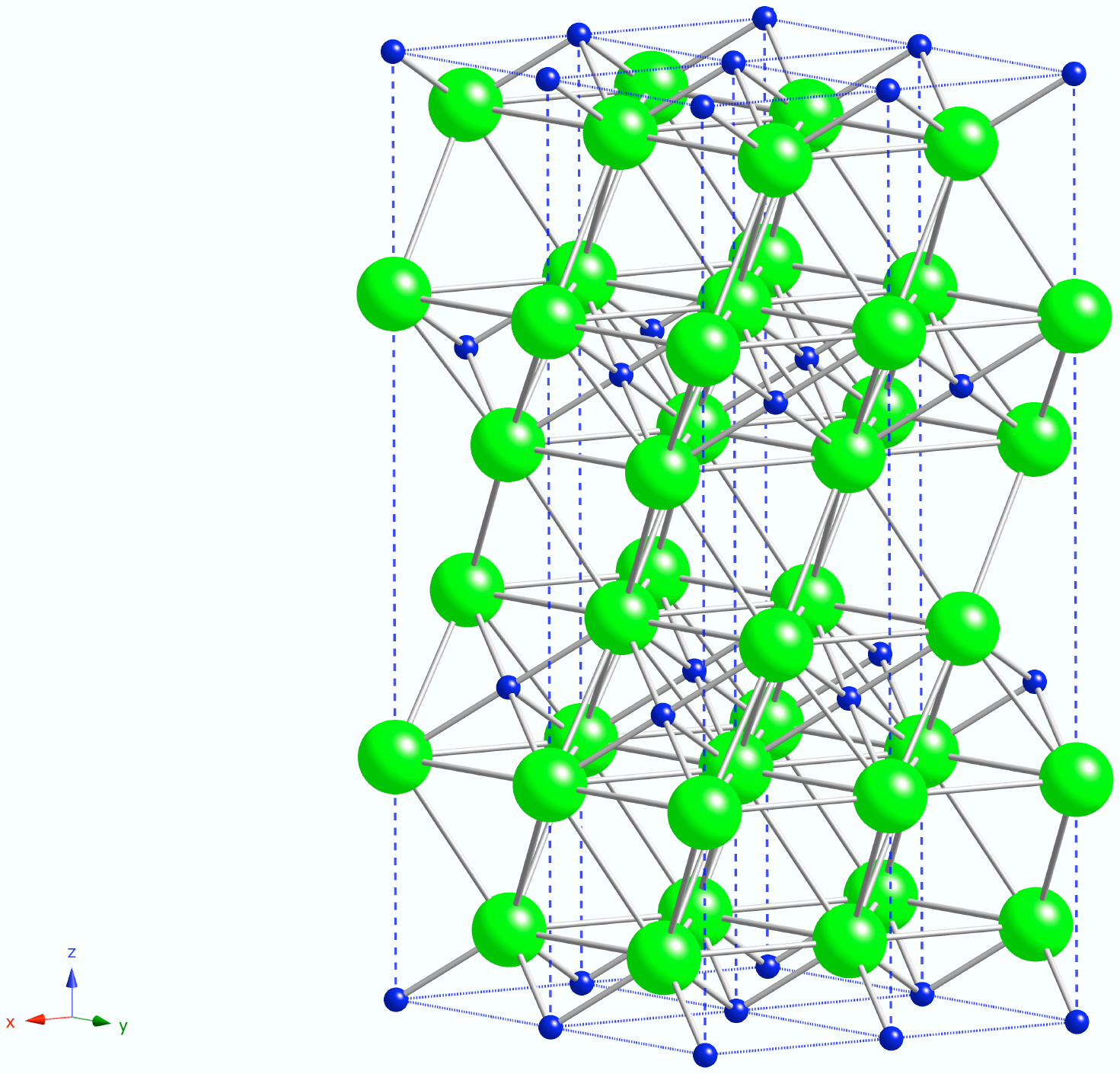}}
\caption {(Color online) Schematic representation of the crystal structure of CoCl$_2$.   Small blue spheres  are Co ions and larger green spheres are Cl ions.          } 
\label{structure}
\end{figure}

The method of investigating hyperfine interaction by high resolution back-scattering neutron spectroscopy was developed by Heidemann \cite{heidemann70}. Heidemann \cite{heidemann70} worked out 
the double differential cross section of this scattering process. The 
process can be summarized as follows: If  neutrons with spin ${\bf s}$ 
are scattered from  nuclei with spins  ${\bf I}$, the probability that 
their spins will be flipped is $2/3$. The nucleus at which the 
neutron is scattered with a spin-flip, changes its magnetic quantum 
number $M$ to $M\pm 1$ due to the conservation of the 
angular momentum. If the nuclear ground state is split up into 
different energy levels $E_{M}$ due to the 
hyperfine magnetic field or an electric quadrupole interaction, then 
the neutron spin-flip produces a change of the ground state energy 
$\Delta E = E_{M} - E_{M\pm 1}$. This energy change is transferred 
to the scattered neutron. The double differential scattering cross section \cite{heidemann70} is 
given by  the following expressions:
\begin{equation}
	 \left(\frac{d^2\sigma}{d\Omega d\omega}\right)_
{inc}^{0}=\overline{(\bar{\alpha^{2}}-{\bar{\alpha}}^{2}+
\frac{1}{3}{\alpha^{\prime}}^{2}I(I+1))}e^{-2W(Q)}
\delta(\hbar\omega),
\label{heidemann01}
\end{equation}
\begin{equation}
\left(\frac{d^2\sigma}{d\Omega d\omega}\right)_
{inc}^{\pm}=
\frac{1}{3}\overline{{\alpha^{\prime}}^{2}I(I+1)}\sqrt{1\pm\frac{\Delta E}{E_{0}}}e^{-2W(Q)}
\delta(\hbar\omega\mp \Delta E)
\label{heidemann02}	 
\end{equation}
where $\alpha$ and $\alpha^{\prime}$ are coherent and spin-incoherent 
scattering lengths, $W(Q)$ is the Debye-Waller factor and $E_{0}$ is 
the incident neutron energy, $\delta$ is the Dirac delta function.  
If the sample contains one type of isotope then 
$\bar{\alpha^{2}}-{\bar{\alpha}}^{2}$ is zero. Also 
$\sqrt{1\pm\frac{\Delta E}{E_{0}}} \approx 1$ because $\Delta E$ is 
usually much less than the incident neutron energy $E_{0}$. In this case 2/3 of 
incoherent scattering will be spin-flip scattering. 
Also one expects a central elastic peak and two inelastic peaks of 
approximately equal intensities. The $^{59}$Co is such a case. 
\begin{figure}
\resizebox{0.35\textwidth}{!}{\includegraphics{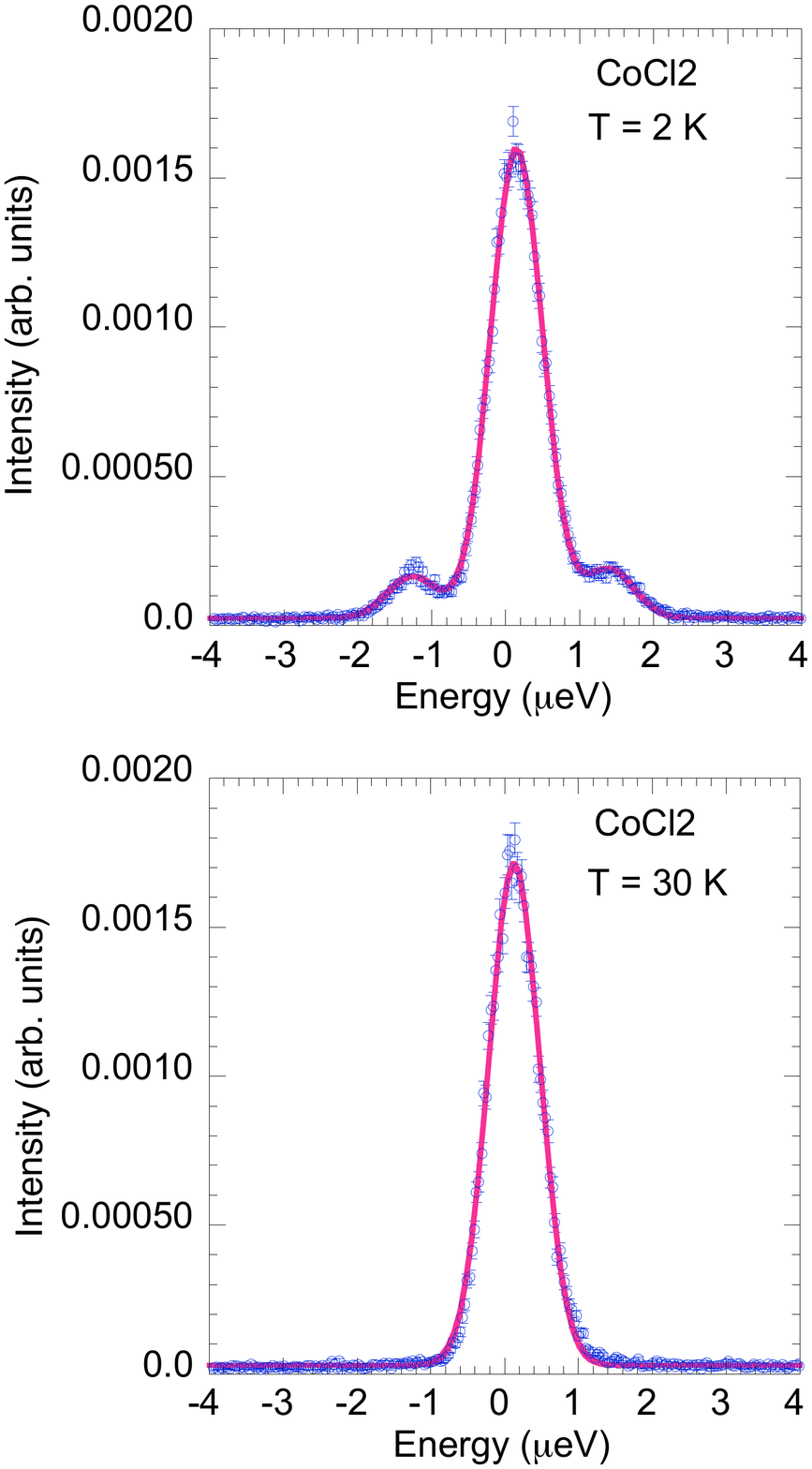}}
\caption {(Color online) Energy spectra of CoCl$_2$ at T = 2 below $T_N\approx 25$ K and at $T = 30$ K above $T_N$ measured on IN16. At $T = 2 $K two inelastic peaks at the energy loss and energy gain sides and an elastic peak in the middile at $E \approx 0$. At T = 30 K the inelastic peaks disappear leaving only the middle elastic peak at E = 0. The continuous curves are fits of the elastic and the inelastic peaks with three Gaussian functions.
            } 
\label{spectra}
\end{figure}

\begin{figure}
\resizebox{0.35\textwidth}{!}{\includegraphics{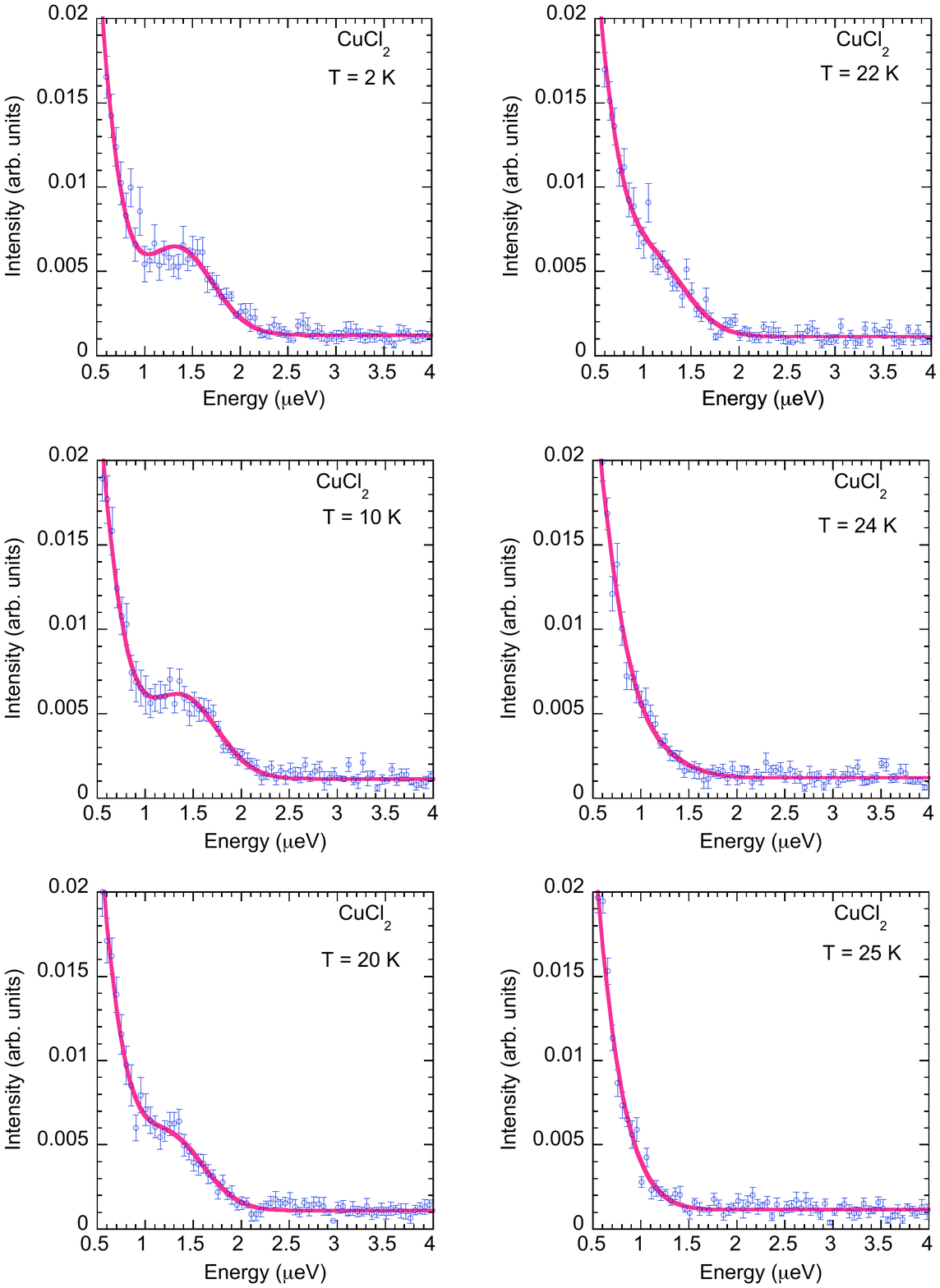}}
\caption {(Color online) Temperature evolution of the inelastic spectra of CoCl$_2$ measured on IN10. Only one inelastic peak has been shown in the spectra.
            } 
\label{spectra}
\end{figure}

\begin{figure}
\resizebox{0.4\textwidth}{!}{\includegraphics{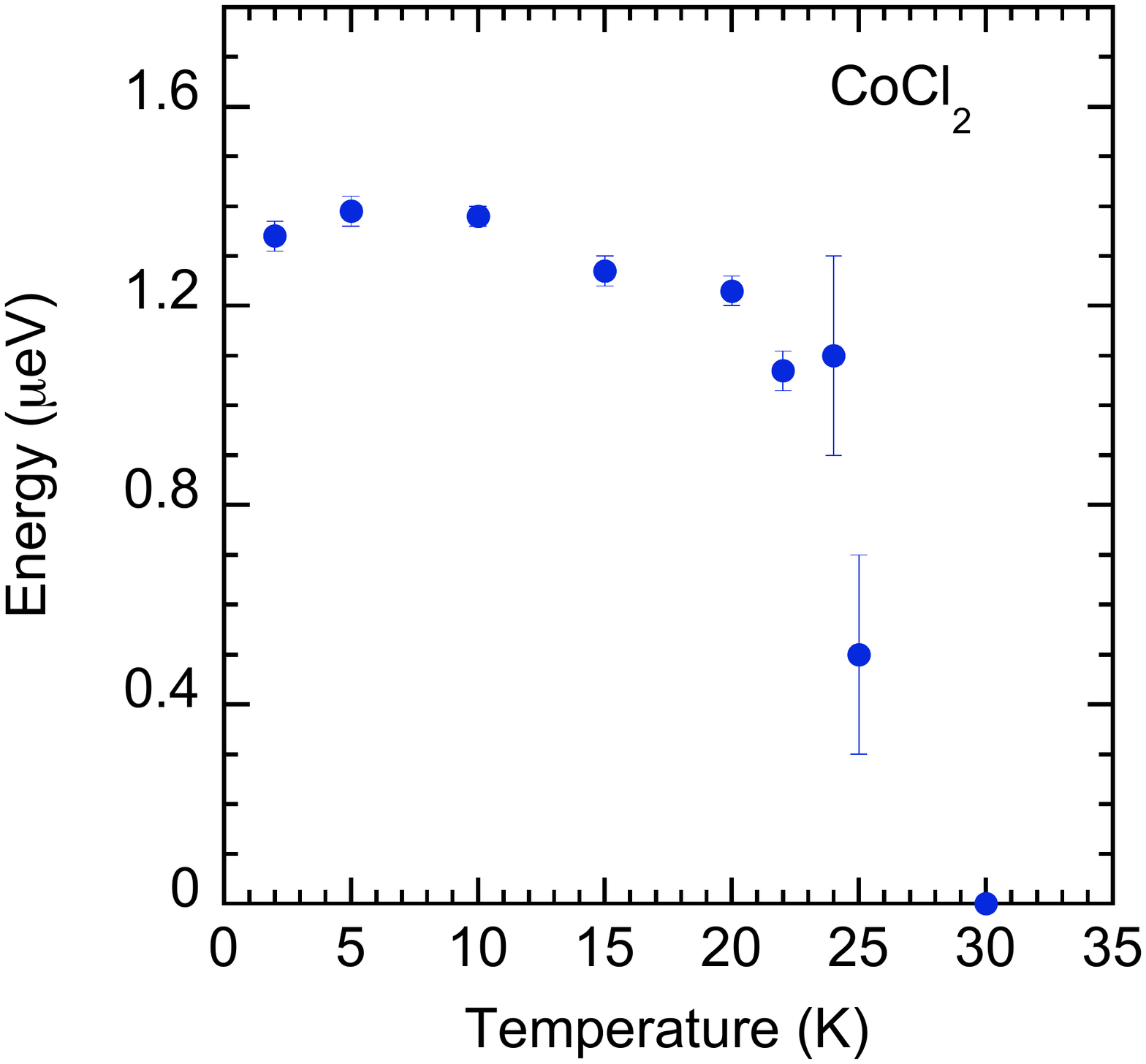}}
\caption {(Color online) Temperature variation of the energy of the inelastic peak 
          of CoCl$_2$. }
\label{E(T)}
\end{figure}

 We investigated previously hyperfine interaction in several Nd compounds  \cite{chatterji00,chatterji02,chatterji04,chatterji04a,chatterji08,chatterji08a,chatterji09f} by high resolution neutron spectroscopy and found that the hyperfine splitting of the Nd nuclear levels is linearly proportional to the ordered electronic magnetic moment of Nd. Our recent investigation on a series of Co compounds \cite{chatterji09c,chatterji09d,chatterji10} showed that this simple relationship is no longer valid for Co compounds presumably due to the unquenched orbital moments in Co-compounds. It is known that in Co metal and also in some Co-compounds the sign of the hyperfine field due the electronic orbital magnetic moment is opposite to that due to the spin moment. The orbital moments in different Co-compounds are different and often unknown. The determination of orbital moment is not easy and involes either polarized neutron diffraction or X-ray magnetic scattering or x-ray magnetic circular dichroism (XRMD) techniques. What we usually know is the total ordered magnetic moment by unpolarised neutron diffraction. So study of hyperfine interaction may also give useful information about the orbital magnetic moment. We therefore studied a series of Co compounds \cite{chatterji09c,chatterji09d,chatterji10} by the high-resolution back-scattering neutron spectroscopy. Here 
we report the results of the study the hyperfine interaction in the layered compound CoCl$_2$. 

Anhydrous CoCl$_2$ crystallizes with the CdCl$_2$ type layered rhombohedral structure (Fig. \ref{structure}) with one molecule per unit cell but a larger hexagonal cell that contains three molecules is normally chosen. In this hexagonal cell with $a = b = 3.553$ {\AA} and $c = 17.359$ {\AA}, $\alpha = \beta = 90 ^\circ$ and $\gamma = 120 ^\circ$ the Co atoms are located at $(000)$ and $(\frac{1}{3},\frac{2}{3},\frac{2}{3})$, and six Cl atoms are located at $(0,0,\pm u), (\frac{1}{3},\frac{2}{3}, \frac{2}{3}\pm u)$, and $(\frac{2}{3},\frac{1}{3}, \frac{1}{3}\pm u)$, where $u = 0.25 \pm 0.01$. The Co atoms are bonded tightly to the Cl atoms in the plane on the either side, but the adjacent planes of halide atoms weakly bounded. Because of these weak bonding between the layers stacking faults are easily produced in these layerd structures. CoCl$_2$ orders below $T_N \approx 25$ K with a simple antiferromagnetic structure in which the Co layers are ferromagnetically arranged individually but stacked antiferromagnetically. In case of CoCl$_2$ the magnetic moment of the Co atoms lie in the $(0001)$ plane along $[2,1,\bar{3},0]$.

  \begin{table}[ht]
\caption{Ordered electronic moment of Co and the energy of Co nuclear spin excitations }
\label{table1}
\begin{center}
\begin{tabular}{lccc} \hline \hline
\emph{Compound} & Moment ($\mu_B$)&\emph{$\Delta E (\mu eV)$}  & \emph{Reference}\\ \hline
CoCl$_2$ & 3.0(6)&1.34(3)&[present work]\\
CoV$_2$O$_6$& 3.5(1)& 1.379(6)&[16]\\
Co$_2$SiO$_4$& 3.61(3)& 1.387(6)&[12]\\
CoF$_2$ & 2.60(4)&0.728(8)&[11]\\
CoO & 3.80(6)&2.05(1)&[10]\\
Co & 1.71 & 0.892(4)&[13,14]\\
Co$_{0.873}$P$_{0.127}$ & 1.35&0.67&[13]\\
Co$_{0.837}$P$_{0.161}$ &1.0& 0.54&[13]\\
Co$_{0.827}$P$_{0.173}$& 1.07&0.56&[13] \\
Co$_{0.82}$P$_{0.18}$ & 0.93&0.49 &[13]\\ 
LaCo$_{13}$&1.58&0.69&[15]\\
LaCo$_{5}$&1.46&0.32&[15]\\
YCo$_{5}$&1.51&0.37&[15]\\
ThCo$_{5}$&1.02&0.31&[15]\\\hline
\end{tabular}
\end{center}
\end{table}
\begin{figure}
\resizebox{0.4\textwidth}{!}{\includegraphics{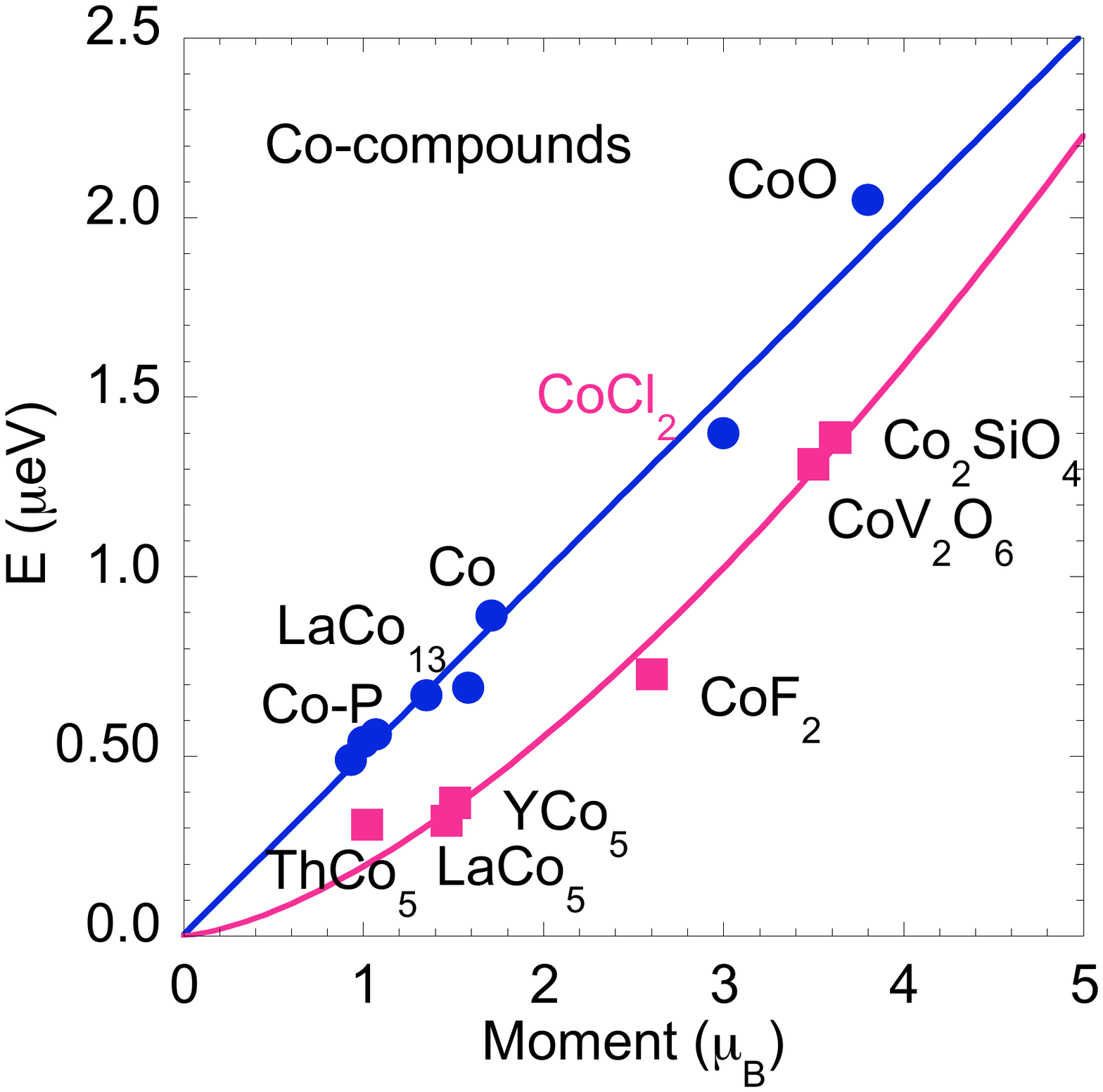}}
\caption {(Color online) Plot of the energy of inelastic signal vs. ordered electronic moment of Co-based materials. The continuous curves are linear and power law fit of the data corresponding to the normal and anomalous compounds.}
\label{cocompounds}
\end{figure}

We performed inelastic neutron scattering experiment on the back-scattering neutron spectrometers IN16 and IN10 of the Institute Laue-Langevin. The neutron wavelength was 6.271 {\AA}. About 5 g of powder CoCl$_2$ sample was placed inside a flat Al sample holder that was fixed to the cold tip of the standard He cryofurnace. Fig. \ref{spectra} shows energy spectra obtained from CoCl$_2$ at several temperatures. We checked carefully the Q dependence of the spectra and found that they were Q independent as expected for hyperfine peaks. So we integrated over all measured Q. At low temperature we see two inelastic peaks on both sides of the central elastic peak at about 1.4 $\mu$eV. The inelastic peaks move towards the central elastic peak at higher temperatures and finally merge into the elastic peak at $T_N \approx 25$ K. We fitted three Gaussian peaks for the elastic and the two inelastic peaks by least squares method. We constrained the two inelastic peaks to have same widths. Fig. \ref{E(T)} shows the temperature variation of the energy of the inelastic peaks. The energy of the inelastic peak decreases continuously at first slowly then close to $T_N \approx 25$ K the energy becomes zero.

  In Table 1 we give the magnetic moments and the energy of low energy nuclear spin excitations in Co and all Co compounds studied so far by high resolution neutron spectroscopy. Hyperfine interaction in Co and amorphus Co-P alloys at room temperature was studied by Heidemann \cite{heidemann75a} by high resolution neutron spectroscopy. Using the same technique we recently studied hyperfine interaction in Co \cite{chatterji10a} in the temperature range $2-1550$ K. Heideman et al. \cite{heidemann75b} studied hyperfine interaction in the intermetallic LaCo$_{13}$, LaCo$_5$, YCo$_5$ and ThCo$_5$. In Fig. \ref{cocompounds} we plot the energy of nuclear spin excitations vs. the ordered magnetic moments of CoV$_2$O$_6$ along with other Co compounds investigated so far. We note that although Co and CoO and amorphous Co-P alloys and also perhaps LaCo$_{13}$ lie on a straight line passing through the origin, several other compounds viz. LaCo$_5$, YCo$_5$, ThCo$_5$ CoF$_2$, CoV$_2$O$_6$ and Co$_2$SiO$_4$ deviate from the linear behaviour. The slope of the linear fit $E = a\mu$ ($\mu$ = magnetic moment) of the
data for  Co, Co-P amorphous alloys, CoO  and LaCo$_{13}$ gives a value of $a=0.51 \pm 0.01 \mu eV$/$\mu_B$. The data for these compounds have been shown by blue circles and the fitted blue straight line. The data corresponding to the other \emph{anomalous compounds} can be fitted by a power law $E=a\mu^n$ with $a = 0.19 \pm 0.03$ and $n= 1.5 \pm 0.1\approx 3/2$ and is shown by the red curve in Fig. \ref{cocompounds}. We ascribe the anomalous behaviour to the presence of unquenched orbital moments in these compounds. The orbital moment 3d transition metal compounds is normally quenched. However Co and also V compounds are known to possess considerable unquenched orbital moments. The hyperfine field due to the unquenched orbital moment can have opposite sign \cite{freeman65} to that due to Fermi contact term and thus can reduce the effective hyperfine field. This is probably the case for the Co compounds, CoF$_2$, Co$_2$SiO$_4$, CoV$_2$O$_6$ and intermetallic compounds LaCo$_{13}$, LaCo$_5$, YCo$_5$ and ThCo$_5$. The hyperfine fields in these compounds are much less than that expected from their moments and therefore deviate from the linear behaviour. This is of course only a qualitative explanation in the absence of any ab-intitio calculations of the orbital moments and hyperfine fields in these compounds.

\end{document}